# Lower Bounds for Exact Model Counting and Applications in Probabilistic Databases[*]

Paul Beame          Jerry Li          Sudeepa Roy          Dan Suciu

Computer Science and Engineering
University of Washington
Seattle, WA 98195
{beame, jerryzli, sudeepa, suciu}@cs.washington.edu

## Abstract

The best current methods for exactly computing the number of satisfying assignments, or the satisfying probability, of Boolean formulas can be seen, either directly or indirectly, as building *decision-DNNF (decision decomposable negation normal form)* representations of the input Boolean formulas. Decision-DNNFs are a special case of *d-DNNFs* where *d* stands for *deterministic*. We show that any decision-DNNF can be converted into an equivalent *FBDD (free binary decision diagram)* – also known as a *read-once branching program (ROBP or 1-BP)* – with only a quasipolynomial increase in representation size in general, and with only a polynomial increase in size in the special case of monotone *k*-DNF formulas. Leveraging known exponential lower bounds for FBDDs, we then obtain similar exponential lower bounds for decision-DNNFs which provide lower bounds for the recent algorithms. We also separate the power of decision-DNNFs from *d*-DNNFs and a generalization of decision-DNNFs known as AND-FBDDs. Finally we show how these imply exponential lower bounds for natural problems associated with probabilistic databases.

## 1   Introduction

Model counting is the problem of computing the number, $\#F$, of satisfying assignments of a Boolean formula $F$. While model counting is hard for $\#\mathsf{P}$, there have been major advances in practical algorithms that compute exact model counts for many relatively complex formulas and, using similar techniques, that compute the probability that Boolean formulas are satisfied, given independent probabilities for their literals.

Modern exact model counting algorithms use a variety of techniques (see [Gomes et al., 2009] for a survey). Many are based on extensions of backtracking search using the *DPLL* family of algorithms [Davis and Putnam, 1960, Davis et al., 1962] that were originally designed for satisfiability search. In the context of model counting (and related problems of exact Bayesian inference) extensions include caching the results of solved sub-problems [Majercik and Littman, 1998], dynamically decomposing residual formulas into components (Relsat [Bayardo et al., 2000]) and caching their counts ([Bacchus et al., 2003]), and applying dynamic component caching together with conflict-directed clause learning (CDCL) to further prune the search (Cachet [Sang et al., 2004] and sharpSAT [Thurley, 2006]).

The other major approach, known as *knowledge compilation*, is to convert the input formula into a representation of the Boolean function that the formula defines and from which the model count can be computed efficiently in the size of the representation [Darwiche, 2001a, Darwiche, 2001b, Huang and Darwiche, 2007, Muise et al., 2012]. Efficiency for knowledge compilation depends both on the size of the representation and the time required to construct it. As noted by (c2d [Huang and Darwiche, 2007] based on component caching) and (Dsharp [Muise et al., 2012] based on sharpSAT), the traces of all the DPLL-based methods yield knowledge compilation algorithms that can produce what are known as *decision-DNNF* representations [Huang and Darwiche, 2005, Huang and Darwiche, 2007], a syntactic subclass of *d*-DNNF representations [Darwiche, 2001b, Darwiche and Marquis, 2002]. Indeed, all the methods for exact model counting surveyed in [Gomes et al., 2009] (and all others of which we are aware) can be converted to knowledge com-pilation algorithms that produce decision-DNNF representations, without any significant increase in their running time.

[*]This work was partially supported by NSF IIS-1115188, IIS-0915054, and CCF-1217099.

In this paper we prove exponential lower bounds on the size of decision-DNNFs for natural classes of formulas. Therefore our results immediately imply exponential lower bounds for all modern exact model counting algorithms. These bounds are unconditional – they do not depend on any unproved complexity-theoretic assumptions. These bounds apply to very simple classes of Boolean formulas, which occur frequently both in uncertainty reasoning, and in probabilistic inference. We also show that our lower bounds extend to the evaluation of the properties of a large class of database queries, which have been studied in the context of probabilistic databases.

We derive our exponential lower bounds by showing how to translate any decision-DNNF to an equivalent *FBDD*, a less powerful representation for Boolean functions. Our translation increases the size by at most a quasipolynomial, and by at most a polynomial in the special case when the Boolean function computed has a monotone $k$-DNF formula. The lower bounds follow from well-established exponential lower bounds for FBDDs. This translation from decision-DNNFs to FBDDs is of independent interest: it is simple, and efficient, in the sense that it can be computed in time linear in the size of the output FBDD.

It is interesting to note that with formula caching, but without dynamic component caching, the trace extensions of DPLL-based searches yield FBDDs rather than decision-DNNFs. Hence, the difference between FBDDs and decision-DNNFs is precisely the ability of the latter to take advantage of decompositions into connected components of subformulas of the formula being represented. Our conversion shows that these connected component decompositions can only provide quasipolynomial improvements in efficiency, or only a polynomial improvement in the case of monotone $k$-DNF formulas.

**Representations** Though closely related, FBDDs and decision-DNNFs originate in completely different approaches for representing (or computing) Boolean functions. FBDDs are special kinds of *binary decision diagrams* [Akers, 1978], also known as *branching programs* [Masek, 1976]. These represent a function using a directed acyclic graph with *decision* nodes, each of which queries a Boolean variable representing an input bit and has 2 out-edges, one labeled 0 and the other 1; it has a single source node, and has sink nodes labeled by output values; the value of the function on an assignment of the Boolean variables is the label of the sink node reached. *Free* binary decision diagrams (FBDDs), also known as *read-once branching programs* (ROBPs), have the property that each input variable is queried at most once on each source-sink path[1]. There are many variants and extensions of these decision-based representations; for an extensive discussion of their theory see the monograph [Wegener, 2000]. These include nondeterministic extensions of FBDDs called *OR-FBDDs*, as well as their corresponding co-nondeterministic extensions called *AND-FBDDs*, which have additional internal AND nodes through which any input can pass – the output value is 1 for an input iff *every* consistent source-sink path leads to a sink labeled 1.

Decision-DNNFs originate in the desire to find restricted forms of Boolean circuits that have better properties for knowledge representation. *Negation normal form (NNF)* circuits are those that have unbounded fan-in AND and OR nodes (gates) with all negations pushed to the input level using De Morgan's laws. Darwiche [Darwiche, 2001a] introduced *decomposable negation normal form (DNNF)* which restricts NNF by requiring that the sub-circuits leading into each AND gate are defined on disjoint sets of variables. He also introduced *d*-DNNFs [Darwiche, 2001a, Darwiche and Marquis, 2002] which have the further restriction that DNNFs are *deterministic*, *i.e.*, the sub-circuits leading into each OR gate never simultaneously evaluate to 1; *d*-DNNFs have the advantage of probabilistic polynomial-time equivalence testing [Huang and Darwiche, 2007]. Most subsequent work has used these *d*-DNNFs. An easy way of ensuring determinism is to have a single variable $x$ that evaluates to 1 on one branch and 0 on the other, so *d*-DNNFs can be produced by the subcircuit $(x \wedge A) \vee (\neg x \wedge B)$, which is equivalent to having decision nodes as above; moreover, the decomposability ensures that $x$ does not appear in either $A$ or $B$. *d*-DNNFs in which all OR nodes are of this form are called decision-DNNFs [Huang and Darwiche, 2005, Huang and Darwiche, 2007]. Virtually all algorithmic methods that use *d*-DNNFs, including those used in exact model counting and Bayesian inference, actually ensure determinism by using decision-DNNFs. Decision-DNNFs have the further advantage of being *syntactically* checkable; by comparison, given a general DNNF, it is not easy to check whether it satisfies the semantic restriction of being a *d*-DNNF.

---

[1]The term *free* contrasts with *ordered* binary decision diagrams (OBDDs) [Bryant, 1986] in which each root-leaf path must query the variables in the same order. For each variable order, minimized OBDDs are canonical representations for Boolean functions, making them extremely useful for a vast number of applications. Unfortunately, OBDDs are often also simply referred to as BDDs, which leads to confusion with the original general model.

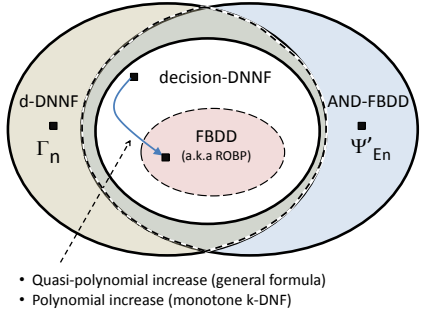

Figure 1: A summary of our contributions (see Section 3). Here, one representation is contained in another if and only if the first can be (locally) translated into the second with at most a polynomial increase in size.

It is immediate that one can get a completely equivalent representation to the above definition by using a decision node on $x$ in place of each OR of ANDs involving $x$, and in place of each leaf variable or its negation; the decomposability property ensures that no root-leaf path in the circuit queries the same variable more than once. Clearly these form a special subclass of the AND-FBDDs discussed above, in which each AND is required to have the decomposability property that the different branches below each AND node query disjoint sets of variables. Though formally there are insignificant syntactic differences between the definitions, we will use the term decision-DNNFs to refer to these *decomposable* AND-FBDDs.

Two other consequences of our simulation of decision-DNNFs by FBDDs are provable exponential separations between the representational power of decision-DNNFs and that of either $d$-DNNFs or AND-FBDDs. There are two functions, involving simple tests on the rows and columns of Boolean matrices, that require exponential size FBDDs but have linear size representations as AND-FBDDs and $d$-DNNFs respectively (cf. Thms 10.3.8, 10.4.7. in [Wegener, 2000]); our simulation shows that these lower bounds carry over to decision-DNNFs, yielding the claimed separations. A comparison of these representations in terms of their succinctness as well as a summary of our contributions in this paper are given in Figure 1[2].

**Probabilistic Databases** These databases annotate each tuple with a probability of being true [Suciu et al., 2011]. Query evaluation on probabilistic databases reduces to the problem of computing the probability of a positive, $k$-DNF Boolean formula, where the number of Boolean variables in each term is bounded by $k$, which is fixed by the query, while the size of the formula grows polynomially in the size of the database. Our results immediately imply that, when applied to such formulas, decision-DNNFs are only polynomially more concise than FBDDs. By combining this with previously known results, we describe a class of queries such that any query in this class generates Boolean formulas requiring decision-DNNFs of exponential size, thus implying that none of the recent evaluation algorithms that either explicitly or implicitly yield decision-DNNFs can compute these queries efficiently. Although the exponential lower bounds we derive for decision-DNNFs are not the first – there are a small number of exponential lower bounds known even for unrestricted AND-FBDDs [Wegener, 2000], which therefore also apply for decision-DNNFs – none of these apply to the kinds of simple structured properties that show up in probabilistic databases that we are able to analyze.

**Compilation** As noted above, the size of the decision-DNNF required is not the only source of complexity in exact model counting. The other source is the search or compilation process itself – the time required to produce a decision-DNNF from an input Boolean formula which may greatly exceed the size of the representation. A particularly striking case where this is an issue is that of an unsatisfiable Boolean formula for which the function evaluates to the constant 0 and hence the decision-DNNF is of size 1. Determining this fact may take exponential time. Indeed, DPLL with caching and conflict-directed clause learning is a special case of resolution theorem proving [Beame et al., 2004]. There are large numbers of unsatisfiable formulas for which exponential lower bounds are known for every resolution refutation (see, *e.g.*, [Ben-Sasson and Wigderson, 2001]) and hence this compilation process must be exponential for such formulas[3]. The same issues can arise in ruling out parts of the space of assignments for satisfiable formulas. However, we do not know of any lower bounds for this excess compilation time that directly apply to the kinds of simple highly satisfiable instances that we discuss in this paper.

The rest of the paper is organized as follows. In Section 2 we review FBDDs and decision-DNNFs. Section 3 presents our two main results: a general transformation of a decision-DNNF into an equivalent FBDD, with only a quasipolynomial increase in size in general, and only a polynomial increase in size for monotone $k$-DNF formulas. We prove these results in Section 4 and Section 5. In Section 6 we discuss the implications of this transformation for evaluating queries in probabilistic databases. We conclude in Section 7.

---

[2]It is open whether the region is empty if no black square is shown (also indicated by dotted borders).

[3]DPLL with formula caching, but not clause learning, can be simulated by even simpler *regular* resolution, though in general it is not quite as powerful as regular resolution [Beame et al., 2010].

## 2 FBDDs and Decision-DNNFs

**FBDDs.** An FBDD is a rooted directed acyclic graph (DAG) $\mathcal{F}$, with two kinds of nodes: *decision nodes*, each labeled by a Boolean variable $X$ and two outgoing edges labeled 0 and 1, and *sink nodes* labeled 0 and 1. Every path from the root to some leaf node may test a Boolean variable $X$ at most once. The size of the FBDD is the number of its nodes. We denote the sub-DAG of $\mathcal{F}$ rooted at an internal node $u$ by $\mathcal{F}_u$ which computes a Boolean function $\Phi_u$; $\mathcal{F}$ computes $\Phi_r$ where $r$ is the root. For a node $u$ labeled $X$ with 0- and 1-children $u_0$ and $u_1$, $\Phi_u = (\neg X)\Phi_{u_0} \vee X\Phi_{u_1}$. The probability of $\Phi_r$ can be computed in linear time in the size of the FBDD using a simple dynamic program: $\Pr[\Phi_u] = (1 - p(X))\Pr[\Phi_{u_0}] + p(X)\Pr[\Phi_{u_1}]$.

**Decision-DNNFs** As noted in the introduction, we choose to define decision-DNNFs as a sub-class of AND-FBDDs. An AND-FBDD [Wegener, 2000] is an FBDD with an additional kind of nodes, called AND-nodes; the function associated to an AND-node $u$ with children $u_1, \ldots, u_r$ is $\Phi_u = \Phi_{u_1} \wedge \ldots \wedge \Phi_{u_r}$. A decision-DNNF, $\mathcal{D}$, is an AND-FBDD satisfying the additional restriction that for any AND-node $u$ and distinct children $u_i, u_j$ of $u$, the sub-DAGS $\mathcal{D}_{u_i}$ and $\mathcal{D}_{u_j}$ do not mention any common Boolean variable $X$.

For the rest of the paper we make two assumptions about decision-DNNFs. First, every AND-node has exactly 2 children, and as a consequence every internal node $u$ has exactly two children $v_1, v_2$, called the *left* and *right child* respectively; second, that every 1-sink node has at most one incoming edge. Both assumptions are easily enforced by at most a quadratic increase in the number of nodes in the decision-DNNF.

## 3 Main Results

In this section we state our two main results and show several applications. We first need some notation. For each node $u$ of a decision-DNNF $\mathcal{D}$, let $M_u$ be the number of AND-nodes in the subgraph $\mathcal{D}_u$. If $u$ is an AND-node, then we have $M_u = 1 + M_{v_1} + M_{v_2}$, because, by definition, the two DAGs $\mathcal{D}_{v_1}$ and $\mathcal{D}_{v_2}$ are disjoint; we will always assume that $M_{v_1} \leq M_{v_2}$ (otherwise we swap the two children of the AND-node $u$), and this implies that $M_u \geq 2M_{v_1} + 1$. We classify the edges of the decision-DNNF into three categories: $(u, v)$ is a *light edge* if $u$ is an AND-node and $v$ its first child; $(u, v)$ is a *heavy edge* if $u$ is an AND-node and $v$ is a its second child; and $(u, v)$ is a *neutral edge* if $u$ is a decision node. We always have $M_u \geq M_v$, while for a light edge we have $M_u \geq 2M_v + 1$.

Let $\mathcal{D}$ be a decision-DNNF, $N$ the total number of nodes in $\mathcal{D}$, $M$ the number of AND-nodes, and $L$ the maximum number of light edges on any path from the root node to some leaf node. Our first main result is:

**Theorem 3.1.** *For any decision-DNNF $\mathcal{D}$ there exists an equivalent FBDD $\mathcal{F}$ computing the same formula as $\mathcal{D}$, with at most $NM^L$ nodes. Moreover, given $\mathcal{D}$, $\mathcal{F}$ can be constructed in time $O(NM^L)$.*

We give the proof in Section 4. We next show that the bound $NM^L$ is quasipolynomial in $N$.

**Corollary 3.2.** *For any decision-DNNF $\mathcal{D}$ with $N$ nodes there exists an equivalent FBDD $\mathcal{F}$ with at most $N2^{\log^2 N}$ nodes.*

*Proof.* Consider any path in $\mathcal{D}$ with $L$ light edges, $(u_1, v_1), (u_2, v_2), \ldots, (u_L, v_L)$. We have $M_{u_i} \geq 2M_{v_i} + 1$ and $M_{v_i} \geq M_{u_{i+1}}$ for all $i$, and we also have $M \geq M_{u_1}$ and $M_{v_L} \geq 0$, which implies $M \geq 2^L - 1$ (by induction on $L$). Therefore, $2^L \leq M + 1 \leq N$ (because $\mathcal{D}$ has at least one node that is not an AND-node), and $NM^L = N2^{L \log M} \leq N2^{\log^2 N}$, proving the claim. □

Our second main result concerns monotone $k$-DNF Boolean formulas, which have applications to probabilistic databases, as we explain in Section 6. We show that in this case any decision-DNNF can be converted into an equivalent FBDD with only a polynomial increase in size. This results from the following lemma, whose proof we give in Section 5:

**Lemma 3.3.** *If a decision-DNNF $\mathcal{D}$ computes a monotone $k$-DNF Boolean formula then every path in $\mathcal{D}$ has at most $k - 1$ AND-nodes.*

Therefore, $L \leq k - 1$, and Theorem 3.1 implies:

**Theorem 3.4.** *For any decision-DNNF $\mathcal{D}$ with $N$ nodes that computes a monotone $k$-DNF Boolean formula then there exists an equivalent FBDD $\mathcal{F}$ with at most $N^k$ nodes.*

We give now several applications of our main results.

**Lower Bounds for DPLL-based Algorithm** We give an explicit Boolean formula on which every DPLL-based algorithm whose trace is a decision-DNNF takes exponential time. We use the following formula introduced by Bollig and Wegener [Bollig and Wegener, 1998]. For any set $E \subseteq [n] \times [n]$ define $\Psi_E = \bigvee_{(i,j) \in E} X_i Y_j$, where $X_1, \ldots, X_n, Y_1, \ldots, Y_n$ are Boolean variables. Let $n = p^2$ where $p$ is a prime number; then each number $0 \leq i < n$ can be uniquely written as $i = a + bp$ where $0 \leq a, b < p$. Define $E_n = \{(i+1, j+1) \mid i = a + bp, j = c + dp, c \equiv (a + bd) \mod p\}$. Then:

**Theorem 3.5.** *[Bollig and Wegener, 1998, Th.3.1] Any FBDD for $\Psi_{E_n}$ has $2^{\Omega(\sqrt{n})}$ nodes.*

Consider the formula $\Phi_n = \bigvee_{1 \leq i,j \leq n} X_i Z_{ij} Y_j$. Any FBDD for $\Phi_n$ has size $2^{\Omega(\sqrt{n})}$, because it can be converted into an FBDD for $\Psi_{E_n}$ by setting $Z_{ij} = 1$ or

$Z_{ij} = 0$, depending on whether $(i,j)$ is in $E_n$ or not. Both $\Psi_{E_n}$ and $\Phi_n$ are monotone, and 2-DNF and 3-DNF respectively, therefore, by Theorem 3.4:

**Corollary 3.6.** *Any decision-DNNF for either $\Psi_{E_n}$ or $\Phi_n$ has $2^{\Omega(\sqrt{n})}$ nodes.*

In particular, any DPLL-based algorithm whose trace is a decision-DNNF will take exponential time on the formulas $\Psi_{E_n}$ and $\Phi_n$.

**Separating decision-DNNFs from AND-FBDDs** We show that decision-DNNFs are strictly weaker than AND-FBDDs. Define $\Psi'_{E_n} = \bigwedge_{(i,j) \in E_n} (X_i \vee Y_j)$, the CNF expression that is the dual of $\Psi_{E_n}$. Since $\Psi'_{E_n}$ is a CNF formula, it admits an AND-FBDD with at most $n^2$ nodes (since $|E_n| \leq n^2$). On the other hand, we show that any decision-DNNF must have $\Omega(2^{n^{1/4}})$ nodes. Indeed, Theorem 3.5 implies that any FBDD for $\Psi'_{E_n}$ has $2^{\Omega(\sqrt{n})}$ nodes. Consider some decision-DNNF $\mathcal{D}$ for $\Psi'_{E_n}$ having $N$ nodes. By Corollary 3.2 we obtain an FBDD $\mathcal{F}$ of size $2^{\log^2 N + \log N}$, which must be $2^{\Omega(\sqrt{n})}$; thus $\log^2 N = \Omega(\sqrt{n})$, hence $\log N = \Omega(n^{1/4})$, and $N = 2^{\Omega(n^{1/4})}$. We have shown:

**Corollary 3.7.** *Decision-DNNFs are exponentially less concise than AND-FBDDs.*

**Separating decision-DNNFs from $d$-DNNFs** Define $\Gamma_n$ on the matrix of variables $X_{ij}$ for $i,j \in [n]$ by $\Gamma_n(X) = f_n(X) \vee g_n(X)$ where $f_n$ is 1 if and only if the parity of all the variables is even and the matrix has an all-1 row and $g_n$ is 1 if and only if the parity of all the variables is odd and the matrix has an all-1 column. Wegener showed (cf. Theorem 10.4.7. in [Wegener, 2000]) that any FBDD for $\Gamma_n$ has $2^{\Omega(n)}$ nodes (therefore, every decision-DNNF requires $2^{\Omega(n^{1/2})}$ nodes). $\Gamma_n$ can also be computed by an $O(n^2)$ size $d$-DNNF, because both $f_n$ and $g_n$ can be computed by $O(n^2)$ size OBDDs, and $f_n \wedge g_n \equiv \texttt{false}$. Hence:

**Corollary 3.8.** *Decision-DNNFs are exponentially less concise than d-DNNFs.*

## 4 Decision-DNNF to FBDD

In this section we prove Theorem 3.1 by describing a construction to convert a decision-DNNF $\mathcal{D}$ to an FBDD $\mathcal{F}$.

### 4.1 Main Ideas

To construct $\mathcal{F}$ we must remove all AND-nodes in $\mathcal{D}$ and replace them with decision nodes. An AND node has two children, $v_1, v_2$; we need to replace this node with an FBDD for the expression $\Phi_{v_1} \wedge \Phi_{v_2}$. Assume that $u$ is the only AND-node in $\mathcal{D}$; then both $\mathcal{D}_{v_1}$ and $\mathcal{D}_{v_2}$ are already FBDDs, and Figure 2(a): stack $\mathcal{D}_{v_1}$ over $\mathcal{D}_{v_2}$, and redirect all 1-sink nodes in $\mathcal{D}_{v_1}$ to the root of $\mathcal{D}_{v_2}$. Clearly this computes the same AND function; moreover it is a correct FBDD because $\mathcal{D}_{v_1}, \mathcal{D}_{v_2}$ do not have any common variable. If this construction worked in general, then the entire decision-DNNF would be converted into an FBDD of exactly the same size.

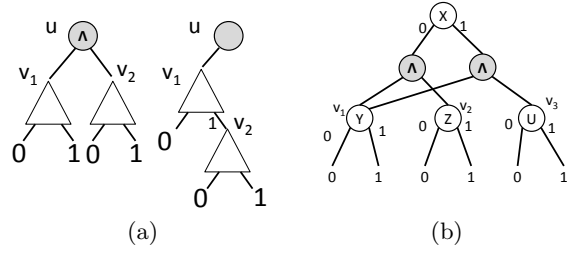

Figure 2: (a) Basic construction for converting a decision-DNNF into an FBDD (b) where it fails.

But, in general, this simple idea fails, as can be seen on the simple decision-DNNF in Figure 2(b) (it computes $(\neg X)YZ \vee XYU$). To compute the first AND node we need to stack $\mathcal{D}_{v_1}$ over $\mathcal{D}_{v_2}$, and to compute the second AND node we need to stack $\mathcal{D}_{v_1}$ over $\mathcal{D}_{v_3}$: this creates a conflict for redirecting the 1-sink node in $\mathcal{D}_{v_1}$ to $v_2$ or to $v_3$[4]. To get around that, we use two ideas. The first idea is to make copies of some subgraphs. For example if we make two copies of $\mathcal{D}_{v_1}$, call them $\mathcal{D}_{v_1}$ and $\mathcal{D}'_{v_1}$, then we can compute the first AND-node by stacking $\mathcal{D}_{v_1}$ over $\mathcal{D}_{v_2}$, and compute the second AND-node by stacking $\mathcal{D}'_{v_1}$ over $\mathcal{D}_{v_3}$ and the conflict is resolved. The second idea is to reorder the children of the AND-nodes to limit the exponential blowup due to copying. We present the details next.

### 4.2 The Construction of $\mathcal{F}$

Fix the decision-DNNF $\mathcal{D}$. Let $u$ denote a node in $\mathcal{D}$ and $P$ denote a path from the root to $u$. Let $s(P)$ be the set of light edges on the path $P$, and let $S(u)$ consist of the sets $s(P)$ for all paths from the root to $u$, formally:

$$s(P) = \{(v,w) \mid (v,w) \text{ is a light edge in } P\}$$
$$S(u) = \{s(P) \mid P \text{ is a path from the root to } u\}$$

We consider the light edges in a set $s = s(P)$ ordered by their occurrences in $P$ (from the root to $u$). This order is independent of $P$: if $s = s(P) = s(P')$ then the light edges occur in the same order on the paths $P$ and $P'$ (since $\mathcal{D}$ is acyclic).

We will convert $\mathcal{D}$ into an FBDD $\mathcal{F}$ with *no-op nodes*, unlabeled nodes having only one outgoing edge. Any FBDD with no-op nodes is easily transformed into a standard FBDD by removing the no-op nodes and redirecting all incoming edges to its unique child.

---

[4] In this particular example one could stack $\mathcal{D}_{v_2}$ and $\mathcal{D}_{v_3}$ over $\mathcal{D}_{v_1}$ and avoid the conflict; but, in general, $\mathcal{D}_{v_2}, \mathcal{D}_{v_3}$ may have conflicts with other subgraphs.

We define $\mathcal{F}$ formally. Its nodes are pairs $(u, s)$ where $u$ is a node in $\mathcal{D}$ and $s \in S(u)$. The root node is $(\texttt{root}(\mathcal{D}), \emptyset)$. The edges in $\mathcal{F}$ are of three types:
**Type 1:** For each light edge $e = (u, v)$ in $\mathcal{D}$ and every $s \in S(u)$, add the edge $((u, s), (v, s \cup \{e\}))$ to $\mathcal{F}$,
**Type 2:** For every neutral edge $(u, v)$ in $\mathcal{D}$ and every $s \in S(u)$ add the edge $((u, s), (v, s))$ to $\mathcal{F}$,
**Type 3:** For every heavy edge $(u, v_2)$, let $e = (u, v_1)$ be the corresponding light sibling edge. Then, for every $s \in S(u)$, add all edges of the form $((w, s \cup \{e\}), (v_2, s))$, where $w$ is a 1-sink node in $\mathcal{D}_{v_1}$, $v_2$ is the heavy child of $u$, $s \cup \{e\} \in S(w)$, and $s \in S(v_2)$.

Finally, we label every node $u' = (u, s)$ in $\mathcal{F}$, as follows: (1) If $u$ is a decision node in $\mathcal{D}$ that tests the variable $X$, then $u'$ is a decision node in $\mathcal{F}$ testing the same variable $X$, (2) If $u$ is an AND-node, then $u'$ is a no-op node, (3) If $u$ is a 0-sink node, then $u'$ is a 0-sink node, (4) If $u$ is a 1-sink node, then: if $s = \emptyset$ then $u'$ is a 1-sink node, otherwise it is a no-op node.

This completes our description of $\mathcal{F}$. The intuition behind it is that, for every AND node, we make a fresh copy of its left child. To illustrate this, suppose $\mathcal{D}$ has a single AND-node $u$ with two children $v_1, v_2$, and let $e = (u, v_1)$ be the light edge. Suppose there is a second, neutral edge into $v_1$, say $(z, v_1)$. Then $\mathcal{F}$ contains two copies of the subgraph $\mathcal{D}_{v_1}$, one with nodes labeled $(w, \{e\})$, and the other with nodes labeled $(w, \emptyset)$. Any 1-sink node in the first copy becomes a no-op node in $\mathcal{F}$ and is connected to $v_2$, similarly to Figure 2(a); the same 1-sink node in the second copy remain 1-sink nodes. This copying process is repeated in $\mathcal{D}_{v_1}$.

### 4.3 Proof of Theorem 3.1

Theorem 3.1 follows from the following three lemmas:

**Lemma 4.1.** $\mathcal{F}$ has at most $NM^L$ nodes.

**Lemma 4.2.** $\mathcal{F}$ is a correct FBDD with no-op nodes.

**Lemma 4.3.** $\mathcal{F}$ computes the same function as $\mathcal{D}$.

*Proof of Lemma 4.1.* The nodes of $\mathcal{F}$ have the form $(u, s)$. There are $N$ possible choices for the node $u$, and at most $M^L$ possible choices for the set $s$, because $|s| \leq L$ (since every path has $\leq L$ light edges), and $M$ is the number of light edges. □

*Proof of Lemma 4.2.* We need to prove three properties of $\mathcal{F}$: that $\mathcal{F}$ is a DAG, that every path in $\mathcal{F}$ reads each variable only once, and that all its nodes are labeled consistently with the number of their children (e.g., a no-op has one child). The first two properties follow from the following claim:

CLAIM: If $u$ is a decision node in $\mathcal{D}$ labeled with a variable $X$, and there exists a non-trivial path (with at least one edge) between the nodes $(u, s)$ $(v, s')$ in $\mathcal{F}$, then the variable $X$ does not occur in $\mathcal{D}_v$.

Indeed, the claim implies that $\mathcal{F}$ is acyclic, because any cycle in $\mathcal{F}$ implies a non-trivial path from some node $(u, s)$ to itself, and obviously $X \in \mathcal{D}_u$, contradicting the claim. It also implies that every path in $\mathcal{F}$ is read-once: if a path tests a variable $X$ twice, once at $(u, s)$ and once at $(u_1, s_1)$, then $X \in \mathcal{D}_{u_1}$, contradicting the claim. It remains to prove the claim.

Suppose to the contrary that there exists a node $(u, s)$ such that $u$ is labeled with $X$ and there exists a path from $(u, s)$ to $(v, s')$ in $\mathcal{F}$ such that $X$ occurs in $\mathcal{D}_v$. Choose $v$ such that $\mathcal{D}_v$ is maximal; i.e., there is no path from $(u, s)$ to some $(v', s'')$ such that $\mathcal{D}_v \subset \mathcal{D}_{v'}$ (in the latter case replace $v$ with $v'$: we still have that $X$ occurs in $\mathcal{D}_{v'}$). Consider the last edge on the path from $(u, s)$ to $(v, s')$ in $\mathcal{F}$:

$$(u, s), \ldots, (w, s''), (v, s') \tag{1}$$

Observe that $(w, v)$ is not an edge in $\mathcal{D}$ since $\mathcal{D}_v$ is maximal and since $(u, v)$ is not an edge in $\mathcal{D}$ by the read-once property of $\mathcal{D}$; therefore, the edge from $(w, s'')$ to $(v, s')$ is of Type 3. Thus, there exists an AND-node $z$ with children $v_1, v$, and our last edge is of the form $(w, s' \cup \{e\}), (v, s')$, where $e = (z, v_1)$ the light edge of $z$. We claim that $e \notin s$; i.e., it is not present at the beginning of the path in (1). If $e \in s$ then, since $s \in S(u)$, we have $u$, which queries $X$, in $D_{v_1}$. Together with the assumption that some node in $D_v$ queries $X$, we see that descendants of the two children $v_1, v$ of AND-node $z$ query the same variable, contradicting the fact that $\mathcal{D}$ is a decision-DNNF. This proves $e \notin s$. On the other hand, $e \in s''$. Now consider the first node on the path in (1) where $e$ is introduced. It can only be an edge of the form $(z, s_1), (v_1, s_1 \cup \{e\})$. But now we have a path from $(u, s)$ to $(z, s_1)$ with $X \in \mathcal{D}_z \supset \mathcal{D}_v$, contradicting the maximality of $v$. This proves the claim.

Finally, we show that all nodes in $\mathcal{F}$ are consistently labeled, *i.e.* they have the correct arity. To prove this, we only need to show that every no-op node has a single child. There are two cases: the node is $(u, s)$ where $u$ is an AND node in $\mathcal{D}$ (for a Type 1 edge), in which case its single child is $(v_1, s \cup \{(u, v_1)\})$; or the node is $(w, s)$ where $w$ is a 1-sink node and $s \neq \emptyset$ (for a Type 3 edge). In that case, let $e = (z, v)$ be the last edge in $s$: more precisely, if $P$ is any path such that $s = s(P)$, then $e$ is the last light edge on $P$. (This $e$ is well defined: if $s = s(P) = s(P')$ then $P$ and $P'$ have the same sets of light edges, and therefore must traverse them in the same order since $\mathcal{D}$ is a DAG.) Let $v'$ be the right child of $z$; then the only edge from $(w, s)$ goes to $(v', s - \{e\})$. □

Next we prove Lemma 4.3, which completes the proof of Theorem 3.1. To prove this we will use the properties that (a) the value of the function computed by an FBDD on an input assignment is the value of the sink reached on the unique path from the root followed by the input, and (b) the value of the function computed by a decision-DNNF is the logical AND of all of the sink values reachable from the root on that assignment.

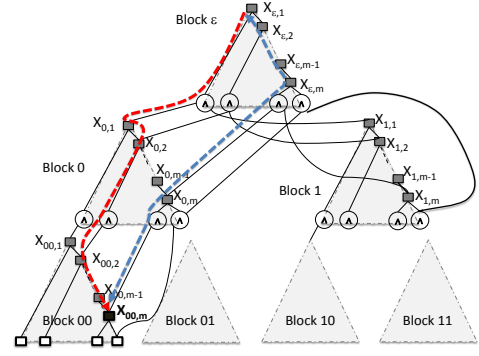

Figure 3: The decision-DNNF $D(p)$, $p = 3$, in Section 4.4. The (red and blue) bold dotted arrows denote two paths from the root to $u = X_{00,m}$. The white boxes at the lowest level denote decision nodes to 0- and 1-sinks.

*Proof of Lemma 4.3.* Let $\Phi_D$ and $\Phi_F$ be the Boolean formulas computed by $\mathcal{D}$ and $\mathcal{F}$ respectively. We show that for any assignment $\theta$ to the Boolean variables, $\Phi_\mathcal{D}[\theta] = 0$ iff $\Phi_\mathcal{F}[\theta] = 0$. For the "if" direction, suppose that $\Phi_\mathcal{F}[\theta] = 0$. Let $P$ be the unique root-sink path in $\mathcal{F}$ consistent with $\theta$, which must reach a 0-sink by assumption. We will show that there exists a path $P'$ in $\mathcal{D}$ from the root to a 0-sink that is consistent with $\theta$. This suffices to prove that $\Phi_\mathcal{D}[\theta] = 0$. First, notice that if $P$ does not contain edges of Type 3, then it automatically also translates into a path leading to a 0-sink in $\mathcal{D}$ and the claim holds. Otherwise, consider an edge of Type 3 from $(w, s \cup \{e\})$ to $(v_2, s)$ such that (i) there exists an AND-node $u$ with children $v_1, v_2$, (ii) $w$ is a descendant of $v_1$, and (iii) $e = (u, v_1)$. Since the edge $e$ must have been introduced along the path, $P$ contains an edge of the form $(u, s'), (v_1, s' \cup \{e\})$. Remove the fragment of $P$ between $(u, s')$ and $(v_2, s)$: this is also a path in $\mathcal{D}$ to a 0-sink (using the original heavy-edge $(u, v_2)$), with one less edge of Type 3, and the claim follows by induction.

For the "only if" part, suppose that $\Phi_\mathcal{D}[\theta] = 0$ and $P'$ is a path in $\mathcal{D}$ from the root to a 0-sink node; as a warm-up, if $P'$ has no heavy edges then it translates immediately into a path in $\mathcal{F}$ to a 0-sink. In general, we proceed as follows. Consider all paths in $\mathcal{D}$ that are consistent with $\theta$ and lead to a 0-sink node. Order them lexicographically as follows: $P'_1 < P'_2$ if, for some $k \geq 1$, $P'_1$ and $P'_2$ agree on the first $k-1$ steps, and at step $k$ $P'_1$ follows the light edge $(u, v_1)$, while $P'_2$ follows the heavy edge $(u, v_2)$ of some AND-node $u$. Let $P'$ be a minimal path under this order. We translate it into a path $P$ in $\mathcal{F}$ iteratively, starting from the root $r$. Suppose we have translated the fragment $r \to u$ of $P'$ into a path $P$ in $\mathcal{F}$: $(r, \emptyset) \to (u, s)$. Consider the next edge $(u, v)$ in $P'$: if it is a light edge $e$ or a neutral edge, we simply extend $P$ with $(v, s \cup \{e\})$ or $(v, s)$ respectively. If $(u, v)$ is a heavy edge, let $(u, v_1)$ be its light sibling, and let $s_1 = s \cup \{e\}$. By the minimality of $P'$, $\Phi_{v_1}[\theta] = 1$ (otherwise we could find a consistent path to a 0-sink in $\mathcal{D}_{v_1}$). We claim that there exists a 1-sink node $w$ in $\mathcal{D}_{v_1}$ s.t. the path $P''$ in $\mathcal{F}_{(v_1, s_1)}$ defined by $\theta$ leads from $(v_1, s_1)$ to $(w, s_1)$: the claim completes the proof of the lemma, because we simply extend $P$ with: $(u, s), (v_1, s_1), P'', (w, s_1), (v, s)$, where the last edge is an edge of Type 3, $(w, s \cup \{e\}), (v, s)$, completing our iterative construction of $P$.

To prove the claim, we apply our decision-DNNF-to-FBDD translation to $\mathcal{D}_{v_1}$, and let $\mathcal{F}_1$ denote the resulting FBDD; by construction, any edge $(z', s'), (z'', s'')$ in $\mathcal{F}_1$ corresponds to an edge $(z', s' \cup s_1), (z'', s'' \cup s_1)$ in $\mathcal{F}$. If $\Phi_{\mathcal{F}_1}$ is the function computed by $\mathcal{F}_1$, then we have already shown that for any $\theta$, $\Phi_{\mathcal{F}_1}[\theta] = 0 \Rightarrow \Phi_{\mathcal{D}_{v_1}}[\theta] = 0$: for our particular $\theta$ we have $\Phi_{\mathcal{D}_{v_1}}[\theta] = \Phi_{v_1}[\theta] = 1$, hence $\Phi_{\mathcal{F}_1}[\theta] = 1$. Therefore, the path defined by $\theta$ in $\mathcal{F}_1$ goes from the root $(v_1, \emptyset)$ to some node $(w, \emptyset)$, where $w$ is a 1-sink node in $\mathcal{D}$; the corresponding path in $\mathcal{F}_{(v_1, s_1)}$ goes from $(v_1, s_1)$ to $(w, s_1)$, proving the claim. □

### 4.4 A Tight Example

We conclude this section by showing that our analysis cannot be tightened to a polynomial bound[5] Fix $M > 0$, and let $m = M^{1/2}$. For each number $p > 0$, the decision-DNNF $\mathcal{D}_p$ given in Figure 3 (for $p = 3$) consists of $m = 2^p - 1$ blocks of size $m$, organized into $p$ levels (0 to $p - 1$). Each block has 2 children to the next level.

A block is identified by $w \in \{0, 1\}^*$, where $|w| \leq p - 1$. Thus, $w = 011$ means "left-right-right" and $w = \epsilon$ means the root block. Each block $w$ has $m + 1$ AND-nodes, $m$ Boolean variables ($X_{w,i}$, where $1 \leq i \leq m$), and $m$ entry points at these $m$ variables. The left (resp. right) child of the $i$-th AND-node in block $w$ points to $X_{w0,i}$ (resp. $X_{w1,i}$), where $1 \leq i \leq m$; The left and the right children of the $(m+1)$-st AND node in block $w$ points to the $(m+1)$-st AND node of blocks $w0$ and $w1$ respectively. Clearly, the total number of AND-nodes in the decision-DNNF is $M = m(m + 1)$.

To obtain a lower bound on the size of the FBDD given by our conversion algorithm, we count the total num-ber of copies $(u, s)$ created for the node $u = X_{00..0,m}$ (i.e., the last decision node in the left-most block at the lowest level), where $s \in S(u)$ is the set of light edges on a path from the root to $u$. For any path $P$ from the root to $u$, let $a_j < m$ be the number of consecutive decision nodes followed by $P$ at the $j$-th level, for $0 \le j \le p-1$; $P$ must take the left (light) branch of the corresponding AND-node at each level $j < m-1$. Note that $\sum_{j=0}^{p-1} a_j = m-1$, any choice of $a_j$'s satisfying this corresponds to a valid path $P$ to $u$, and distinct choices correspond to different sets of light edges. Therefore, $|S(u)|$ is the number of different choices of $a_j$ which is $\binom{m-1+p}{p} \ge \binom{m}{p} \ge (m/p)^p = 2^{p(\log m - \log p)} = 2^{\Omega(\log^2 m)} = 2^{\Omega(\log^2 M)}$ since $p = \Theta(\log m)$ and $M = m(m+1)$.

---
[5]This only applies to our construction. It does not separate FBDDs from decision-DNNFs, since a smaller equivalent FBDD may exist for this decision-DNNF.

## 5 Monotone $k$-DNFs

We prove Lemma 3.3 in this section. Fix a decision-DNNF $\mathcal{D}$ computing a monotone $k$-DNF Boolean function $\Phi$; w.l.o.g. we assume that $\mathcal{D}$ is *non-redundant*: each child of each AND-node in $\mathcal{D}$ computes a non-constant function.

**Proposition 5.1.** $\forall$ *node* $u \in \mathcal{D}$, $\Phi_u$ *is monotone*.

*Proof.* The statement is true for the root node $u$. Suppose that $\Phi_u$ is monotone at some node $u$. If $u$ is a decision node testing the variable $X$ and with children $v_0, v_1$, then both $\Phi_{v_0} = \Phi_u[X=0]$ and $\Phi_{v_1} = \Phi_u[X=1]$ are monotone. If $u$ is an AND-node with children $u_1, u_2$ then $\Phi_u = \Phi_{u_1} \wedge \Phi_{u_2}$ where $\Phi_{u_1}, \Phi_{u_2}$ have disjoint sets of variables, hence they are themselves monotone. The proposition follows by induction. □

In the case of a monotone function $\Phi$, a *prime implicant* is a minimal set of variables whose conjunction implies $\Phi$ and a minimal DNF for $\Phi$ has one term for each of its prime implicants; hence, $\Phi$ can be written as $k$-DNF iff $k$ is the size of its largest prime implicant. If $\theta$ is a partial assignment, then $\Phi[\theta]$ is a $k'$-DNF for some $k' \le k$. Let $A_u$ be the largest number of AND-nodes on any path from the node $u$ to some leaf. The following proposition proves Lemma 3.3:

**Lemma 5.2.** *For every node $u$ with $A_u \ge 1$, if $\Phi_u$ is a monotone $k$-DNF, then $k \ge A_u + 1$.*

*Proof.* The following claim, which we prove by induction on $|A_u|$, suffices to show the lemma: for every node $u$ with $A_u \ge 1$, there exists a partial assignment $\theta$ such that $\Phi_u[\theta]$ is a Boolean formula that is the conjunction of $\ge A_u + 1$ variables.

Observe that it suffices to prove the claim when $u$ is an AND-node, since for any $u'$ with $A_{u'} \ge 1$ that is not an AND-node, there is some AND-node $u$ reachable from $u'$ only via decision nodes (and hence with $A_u = A_{u'}$) and we can obtain the partial assignment $\theta'$ for $u'$ by adding the partial assignment $\sigma$ determined by the path from $u'$ to $u$ to the partial assignment $\theta$ for $u$.

If $u$ is an AND-node with children $v_1, v_2$, then $\Phi_u = \Phi_{v_1} \wedge \Phi_{v_2}$ where $\Phi_{v_1}, \Phi_{v_2}$ do not share any variables. Consider a path starting at $u$ that has $A_u$ AND-nodes and assume w.l.o.g. that it takes the first branch, to $v_1$: thus, $A_u = A_{v_1} + 1$. If $A_{v_1} = 0$ then, since $\mathcal{D}$ is non-redundant, $\Phi_{v_1}$ is non-constant, so there is partial assignment $\theta_1$ such that $I_1 = \Phi_{v_1}[\theta_1]$ is a conjunction of size $\ge 1 = A_{v_1} + 1$. If $A_{v_1} \ge 1$, by the induction hypothesis, there exists a partial assignment $\theta_1$ such that $I_1 = \Phi_{v_1}[\theta_1]$ is a conjunction of size $\ge A_{v_1} + 1$. Since $\mathcal{D}$ is non-redundant, $\Phi_{v_2}$ is non-constant, so there exists a partial assignment $\theta_2$ such that $I_2 = \Phi_{v_2}[\theta_2]$ is a conjunction of size $\ge 1$. Taking $\theta = \theta_1 \cup \theta_2$ and using the disjointness of the variables in $\Phi_{v_1}$ and $\Phi_{v_2}$, we get that $\Phi_u[\theta] = I_1 \wedge I_2$ is a conjunction of size $\ge (A_{v_1} + 1) + 1 = A_u + 1$, proving the claim. □

## 6 Lower Bounds in Probabilistic Databases

We now show an important application of our main result to probabilistic databases. While in knowledge compilation there exists a single complexity parameter, which is the size of the input formula, in databases there are two parameters: the database query, and the database instance. For example, the query may be expressed in a query language, like SQL, and is usually very small (e.g. few lines), while the database instance is very large (e.g. billions of tuples). We are interested here in *data complexity* [Vardi, 1982], where the query is fixed, and the complexity parameter is the size of the database instance. We use Theorem 3.4 to prove an exponential lower bound for the query evaluation problem for every query that is *non-hierarchical*.

We first briefly review the key concepts in probabilistic databases, and refer the reader to [Abiteboul et al., 1995, Suciu et al., 2011] for details.

A relational vocabulary consists of $k$ relation names, $R_1, \ldots, R_k$, where each $R_i$ has an arity $a_i > 0$. A (deterministic) database instance is $D = (A, R_1^D, \ldots, R_k^D)$, were $A$ is a set of constants called the *domain*, and for each $i$, $R_i^D \subseteq A^{a_i}$. Let $n = |A|$ be the size of the domain of the database instance.

A *Boolean query* is a function $Q$ that takes as input a database instance $D$ and returns an output $Q(D) \in \{\texttt{false}, \texttt{true}\}$. A *Boolean conjunctive query* (CQ) is given by an expression of the form $Q = \exists x_1 \ldots \exists x_\ell (P_1 \wedge \ldots \wedge P_m)$, where each $P_k$ is a positive relational atom of the form $R_i(x_{p_1}, \ldots, x_{p_{a_i}})$, with $x_j$ either a variable $\in \{x_1, \ldots, x_\ell\}$ or a constant. A *Boolean Union of Conjunctive Queries* (UCQ) is given by an expression $Q = Q_1 \vee \ldots \vee Q_m$ where each $Q_i$ is a

| Patient P | |
|---|---|
| name | disease |
| Ann | asthma |
| Bob | asthma |
| Carl | flue |

$X_1$
$X_2$
$X_3$

| Friend F | | |
|---|---|---|
| name1 | name2 | |
| Ann | Joe | $Z_{11}$ |
| Ann | Tom | $Z_{12}$ |
| Bob | Tom | $Z_{22}$ |
| Carl | Tom | $Z_{32}$ |

| Smoker S | |
|---|---|
| name | |
| Joe | $Y_1$ |
| Tom | $Y_2$ |

Query $Q = \exists x\, \exists y\, \text{P}(x, \text{'asthma'}) \wedge \text{F}(x,y) \wedge \text{S}(y)$

Lineage expression $\Phi_Q^D = X_1 Z_{11} Y_1 \vee X_1 Z_{12} Y_2 \vee X_2 Z_{22} Y_2$

Figure 4: A database instance, query, and lineage

Boolean conjunctive query. We assume all queries to be minimized (i.e. they do not have redundant atoms, see [Abiteboul et al., 1995]).

Given an instance $D$ and query expression $Q$, the *lineage* $\Phi_Q^D$ is a Boolean formula obtained by grounding the atoms in $Q$ with tuples in $D$; it is similar to *grounding* in knowledge representation [Domingos and Lowd, 2009]. Formally, each tuple $t$ in the database $D$ is associated with unique Boolean variable $X_t$, and the lineage is defined inductively on the structure of $Q$: (1) $\Phi_Q^D = X_t$, if $Q$ is the ground tuple $t$, (2) $\Phi_{Q_1 \wedge Q_2}^D = \Phi_{Q_1}^D \wedge \Phi_{Q_2}^D$, (3) $\Phi_{Q_1 \vee Q_2}^D = \Phi_{Q_1}^D \vee \Phi_{Q_2}^D$, and (4) $\Phi_{\exists x.Q}^D = \bigvee_{a \in A} \Phi_{Q[a/x]}^D$. The lineage is always a monotone $k$-DNF of size $O(n^\ell)$, where $n$ is the domain size and $k, \ell$ are the largest number of atoms, and the largest number of variables in any conjunctive query $Q_i$ of $Q$.

In a *probabilistic database* [Suciu et al., 2011], every tuple $t$ in the database instance is uncertain, and the Boolean variable $X_t$ indicates whether $t$ is present or not. The probability $P(X_t = \text{true})$ is known for every tuple $t$, and is stored in the database as a separate attribute of the tuple. The goal in probabilistic databases is: given a query $Q$ and an input database $D$, compute the probability of its lineage, $P(\Phi_Q^D)$.

**Example 6.1.** *The following example is adapted from [Jha et al., 2010], on a vocabulary with three relations Patient(name, diseases), Friend(name1, name2), Smoker(name) (see Figure 4). Each tuple is associated with a Boolean variable ($X_1, X_2$ etc). The Boolean conjunctive query $Q$ (as well as the lineage $\Phi_Q^D$ for database $D$) returns true if the database instance contains an asthma patient who has a smoker friend. Our goal is to compute $P(\Phi_Q^D)$, given the probabilities of each Boolean variable, when $Q$ is fixed and $D$ is variable.*

**Lemma 6.2.** *Let $h$ be the conjunctive query $\exists x \exists y R(x) \wedge S(x,y) \wedge T(y)$. Then any decision-DNNF for the Boolean formula $\Phi_h^D$ has size $2^{\Omega(\sqrt{n})}$, where $n$ is the size of the domain of $D$.*

*Proof.* For any $n$, let $D$ be the database instance $R^D = [n]$, $S^D = [n] \times [n]$, $T^D = [n]$. Then the lineage $\Phi_h^D$ is exactly the formula $\Phi_n$ of Corollary 3.6, up to variable renaming, and the claim follows. □

Fix a conjunctive query $q = \exists x_1 \ldots \exists x_\ell P_1 \wedge \ldots \wedge P_m$. For each variable $x_j$, let $at(x_j)$ denote the set of atoms $P_i$ that contain the variable $x_j$.

**Definition 6.3.** *[Suciu et al., 2011] The query $q$ is called* hierarchical *if for any two distinct variables $x_i, x_j$, one of the following holds: $at(x_i) \subseteq at(x_j)$, or $at(x_i) \supseteq at(x_j)$, or $at(x_i) \cap at(x_j) = \emptyset$. A Boolean Union of Conjunctive Queries $Q = q_1 \vee \ldots \vee q_k$ is called* hierarchical *if every $q_i$ is hierarchical for $i \in [k]$.*

For example, the query $h$ in Lemma 6.2 is non-hierarchical, because $at(x) = \{R, S\}$, $at(y) = \{S, T\}$, while the query $\exists x.\exists y.R(x) \wedge S(x,y)$ is hierarchical. It is known that, for a non-hierarchical UCQ $Q$, computing the probability of the Boolean formulas $\Phi_Q^D$ is #P-hard [Suciu et al., 2011]. In the full paper we prove:

**Theorem 6.4.** *Consider any Boolean Union of Conjunctive Queries $Q$. If $Q$ is non-hierarchical, then the size of the decision-DNNF for the Boolean functions $\Phi_Q^D$ is $2^{\Omega(\sqrt{n})}$, where $n$ is the size of the domain of $D$.*

The query $Q$ in Example 6.1 is non-hierarchical: $at(x) = \{\text{Patient}, \text{Friend}\}$ and $at(y) = \{\text{Friend}, \text{Smoker}\}$. Therefore, any decision-DNNF computing $Q$ has size $2^{\Omega(\sqrt{n})}$. For another example, consider the following non-hierarchical query that returns true iff the database contains a triangle of friends:

$$q' = \exists x \exists y \exists z (\text{F}(x,y) \wedge \text{F}(y,z) \wedge \text{F}(z,x))$$

Its lineage is $\Delta_n = \bigvee_{i,j,k=1,n} Z_{ij} Z_{jk} Z_{ki}$. By Theorem 6.4, any decision-DNNF for $\Delta_n$ has size $2^{\Omega(\sqrt{n})}$.

## 7 Conclusions and Open Problems

We have proved that any decision-DNNF can be efficiently converted into an equivalent FBDD that is at most quasipolynomially larger, and at most polynomially larger in the case of $k$-DNF formulas. As a consequence, known lower bounds for FBDDs imply lower bounds for decision-DNNFs and thus (a) exponential separations of the representational power of decision-DNNFs from that of both $d$-DNNFs and AND-FBDDs and (b) lower bounds on the running time of any algorithm that, either explicitly or implicitly, produces a decision-DNNF, including the current generation of exact model counting algorithms.

Some natural questions arise: Is there a polynomial simulation of decision-DNNFs by FBDDs for the general case? In particular, is there a polynomial-size FBDD for the example Section 4.4? Is there some other, more powerful syntactic subclass of $d$-DNNFs that is useful for exact model counting? What can be said about the limits of *approximate* model counting [Gomes et al., 2009]?

# References

[Abiteboul et al., 1995] Abiteboul, S., Hull, R., and Vianu, V. (1995). *Foundations of Databases*.

[Akers, 1978] Akers, S. B. (1978). Binary decision diagrams. *IEEE Trans. Computers*, 27(6):509–516.

[Bacchus et al., 2003] Bacchus, F., Dalmao, S., and Pitassi, T. (2003). Algorithms and complexity results for #sat and bayesian inference. In *FOCS*, pages 340–351.

[Bayardo et al., 2000] Bayardo, R. J., Jr., and Pehoushek, J. D. (2000). Counting models using connected components. In *AAAI*, pages 157–162.

[Beame et al., 2010] Beame, P., Impagliazzo, R., Pitassi, T., and Segerlind, N. (2010). Formula caching in dpll. *TOCT*, 1(3).

[Beame et al., 2004] Beame, P., Kautz, H. A., and Sabharwal, A. (2004). Towards understanding and harnessing the potential of clause learning. *J. Artif. Intell. Res. (JAIR)*, 22:319–351.

[Ben-Sasson and Wigderson, 2001] Ben-Sasson, E. and Wigderson, A. (2001). Short proofs are narrow – resolution made simple. *J. ACM*, 48(2):149–169.

[Bollig and Wegener, 1998] Bollig, B. and Wegener, I. (1998). A very simple function that requires exponential size read-once branching programs. *Inf. Process. Lett.*, 66(2):53–57.

[Bryant, 1986] Bryant, R. E. (1986). Graph-based algorithms for boolean function manipulation. *IEEE Trans. Computers*, 35(8):677–691.

[Darwiche, 2001a] Darwiche, A. (2001a). Decomposable negation normal form. *J. ACM*, 48(4):608–647.

[Darwiche, 2001b] Darwiche, A. (2001b). On the tractable counting of theory models and its application to truth maintenance and belief revision. *Journal of Applied Non-Classical Logics*, 11(1-2):11–34.

[Darwiche and Marquis, 2002] Darwiche, A. and Marquis, P. (2002). A knowledge compilation map. *J. Artif. Int. Res.*, 17(1):229–264.

[Davis et al., 1962] Davis, M., Logemann, G., and Loveland, D. (1962). A machine program for theorem-proving. *Commun. ACM*, 5(7):394–397.

[Davis and Putnam, 1960] Davis, M. and Putnam, H. (1960). A computing procedure for quantification theory. *J. ACM*, 7(3):201–215.

[Domingos and Lowd, 2009] Domingos, P. and Lowd, D. (2009). *Markov Logic: An Interface Layer for Artificial Intelligence*.

[Gomes et al., 2009] Gomes, C. P., Sabharwal, A., and Selman, B. (2009). Model counting. In *Handbook of Satisfiability*, pages 633–654.

[Huang and Darwiche, 2005] Huang, J. and Darwiche, A. (2005). Dpll with a trace: From sat to knowledge compilation. In *IJCAI*, pages 156–162.

[Huang and Darwiche, 2007] Huang, J. and Darwiche, A. (2007). The language of search. *JAIR*, 29:191–219.

[Jha et al., 2010] Jha, A. K., Gogate, V., Meliou, A., and Suciu, D. (2010). Lifted inference seen from the other side : The tractable features. In *NIPS*, pages 973–981.

[Majercik and Littman, 1998] Majercik, S. M. and Littman, M. L. (1998). Using caching to solve larger probabilistic planning problems. In *AAAI*, pages 954–959.

[Masek, 1976] Masek, W. J. (1976). *A fast algorithm for the string editing problem and decision graph complexity*. Master's thesis, MIT.

[Muise et al., 2012] Muise, C., McIlraith, S. A., Beck, J. C., and Hsu, E. I. (2012). Dsharp: fast d-dnnf compilation with sharpsat. In *Canadian AI*, pages 356–361.

[Sang et al., 2004] Sang, T., Bacchus, F., Beame, P., Kautz, H. A., and Pitassi, T. (2004). Combining component caching and clause learning for effective model counting. In *SAT*.

[Suciu et al., 2011] Suciu, D., Olteanu, D., Ré, C., and Koch, C. (2011). *Probabilistic Databases*.

[Thurley, 2006] Thurley, M. (2006). sharpsat: counting models with advanced component caching and implicit bcp. In *SAT*, pages 424–429.

[Vardi, 1982] Vardi, M. Y. (1982). The complexity of relational query languages. In *STOC*, pages 137–146.

[Wegener, 2000] Wegener, I. (2000). *Branching programs and binary decision diagrams: theory and applications*.